\RequirePackage{fixltx2e}
\documentclass[aps,pre,preprint,showpacs,showkeys,superscriptaddress,nofootinbib,longbibliography,floatfix]{revtex4-1}
\usepackage[english]{babel}
\usepackage[utf8]{inputenc}
\usepackage[T1]{fontenc}
\usepackage{amsmath,amsfonts,amssymb,graphicx,bbm}
\usepackage{xcolor}
\usepackage{dcolumn}
\usepackage{mathtools}
\usepackage[section]{placeins}
\usepackage{algorithm}  
\usepackage{algpseudocode}

\newcommand{\mathe}{\mathrm{e}}

\usepackage[draft]{minted}
\input{mintedarxiv}

\begin{document}

\title{ Finite symmetries in agent-based epidemic models} 
\author{Gilberto M. Nakamura}
\email{gmnakamura@usp.br}
 \affiliation{Universidade de S\~{a}o Paulo, Faculdade de Filosofia,
   Ciências e Letras de Ribeirão Preto (FFCLRP), Ribeir\~{a}o Preto 14040-901, Brazil}
 \author{Ana Carolina P. Monteiro}
 \email{ana.carolina.monteiro@usp.br}
 \affiliation{Universidade de S\~{a}o Paulo, Faculdade de Filosofia,
   Ciências e Letras de Ribeirão Preto (FFCLRP), Ribeir\~{a}o Preto 14040-901, Brazil}
 \author{George C. Cardoso}
\email{gcc@usp.br}
 \affiliation{Universidade de S\~{a}o Paulo, Faculdade de Filosofia,
   Ciências e Letras de Ribeirão Preto (FFCLRP), Ribeir\~{a}o Preto 14040-901, Brazil}
\author{Alexandre S. Martinez}
\email{asmartinez@ffclrp.usp.br}
 \affiliation{Universidade de S\~{a}o Paulo, Faculdade de Filosofia,
   Ciências e Letras de Ribeirão Preto (FFCLRP), Ribeir\~{a}o Preto 14040-901, Brazil}
\altaffiliation{Instituto Nacional de Ci\^{e}ncia e Tecnologia - Sistemas Complexos (INCT-SC)}

\begin{abstract}
We present an algorithm which explores permutation symmetries
to describe the time evolution of agent-based epidemic models. The
main idea to improve computation times relies on restricting the
stochastic process to one sector of the vector space, labeled by a
single permutation eigenvalue. In this scheme, the transition
matrix reduces to block diagonal form, enhancing computational
performance. 
\end{abstract}

\maketitle

In recent years, the emergence of Zika and Ebola viruses have
attracted much attention from scientific community after
reports of their aggressive effects, microcephaly in newborns
\cite{mlakarNEJM2016} and high mortality rate 
\cite{magangaNEJM2014,ebolaNEJM2015,kerkhoveSciData2015},
respectively. Despite their intrinsic transmission differences, both
viruses spread in a population starting from a single infected
individual based on her geographic 
localization and relationship network. Contact tracing and proper
clinical care planning are key parts of the World Health Organization
(WHO) strategic plan \cite{whoNEJM2016} to mitigate on-going transmissions and
incidence cases, requiring the correct spatiotemporal dissemination of the
disease. This assertion has renewed the interest in agent-based epidemic models
(ABEM).   

\begin{figure}[bh]
  \includegraphics[width=0.4\columnwidth]{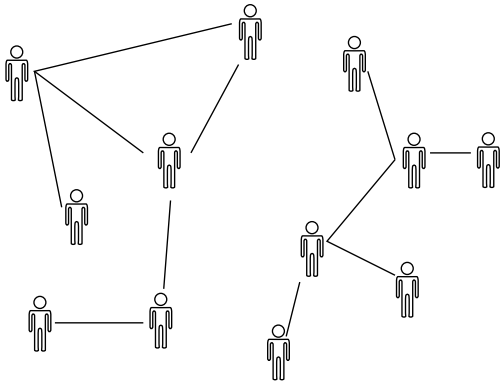}
  \caption{\label{fig_network} Agent network. Agents (vertices) and their
    interconnections (edges) are expressed as a graph. The graph
    representation introduces heterogeneity among the agents, which must
  be accounted for during disease spreading.} 
\end{figure}

ABEM are mathematical models that describe the evolution of infectious
diseases among a finite number $N$ of agents, along time. For that
purpose, agents are labeled using integer numbers $k=0,1,\ldots,N-1$
whereas contacts between agents are mapped \emph{via} an adjacency matrix
$A$. The matrix elements are $A_{i j} = 1$ if the $j$-th agent connects to
the $i$-th agent and vanishes otherwise. Accordingly, the set formed by
agents and their interconnection is expressed as a graph as depicted in
Fig.~\ref{fig_network}. In this way, heterogeneity arises naturally 
since the individuality of agents is taken into account,
distinguishing ABEM from compartmental epidemic models
\cite{keelingJRSoc2005}. 

\begin{figure}
\includegraphics[width=0.25\columnwidth]{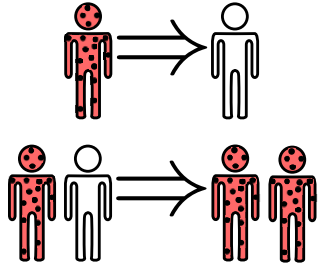}
\caption{\label{fig_events} SIS transition events. Infected agents (red dotted)
  undergo cure events with probability $\gamma$ and change to
  susceptible (empty) health status. Infected agents may also infect
  additional susceptible agents  with probability $\beta$, as long they
  are connected.} 
\end{figure}

The simplest ABEM, the susceptible-infected-susceptible model (SIS),
considers only two health states for agents, infected $\lvert
1 \rangle$ or susceptible $\lvert 0 \rangle$, and the occurrence of the
following events during a time interval $\delta t$
\cite{martinezJStatPhys2011}. An infected agent 
may undergo a cure event and return to susceptible state with probability
$\gamma$; an infected agent may infect a susceptible agent with
transmission probability $\beta$ if and only if both agents are connected;
or remains unchanged, as Fig.~\ref{fig_events}
illustrates. Therefore, the SIS ABEM is inherently a Markov
process. The time interval $\delta t$ is often chosen so that sequential
 cure-cure or transmission-cure events are unlikely within the available
 time window. This is the so-called Poissonian hypothesis
 \cite{satorrasRevModPhys2015}.   

Following Ref.~\cite{nakamuraArxiv2016}, any configuration of $N$ agents is
obtained by direct composition of individual agent states. Let $\mu$ be
an integer that labels the $\mu$-th configuration so that 
\begin{equation}
\lvert \mu \rangle \equiv \lvert n_{N-1}  \cdots n_1\, n_0 \rangle, 
\label{eq:integer_representation}
\end{equation}
with $n_k= 0, 1$ and $\mu=n_{N-1} 2^{N-1} +\cdots+ n_0 2^0$. A simple
example for $N=4$ is $\lvert 8 \rangle \equiv \lvert 1 0 0 0 \rangle$,
which represents the configuration where only the third agent is
currently infected. From this scheme, it is already clear that there
exists $2^N$ configurations in total since there are two available
states for each agent. In what follows, we employ the notation: Latin 
indices enumerate agents $0,1,\ldots,N-1$ while Greek indices enumerate
configuration states $0,1,\ldots,2^N-1$. 

Let $\lvert \pi(t) \rangle$ be the probability vector and $\pi_{\mu}(t) =
\langle \mu \vert \pi(t) \rangle$ the probability of observing the
configuration $\lvert \mu \rangle$ at time $t$
\cite{alcarazPhysRevE2008,alcarazAnnPhys1994}. The master equation for the general Markov
process reads
\begin{equation}
\frac{d}{d t} \pi_{\mu}(t) = - \sum_{\nu}H_{\mu\nu}\pi_{\nu}(t),
\end{equation}
$\hat{H} = (\mathbbm{1} - \hat{T})/\delta t$ is the time generator
whereas $\hat{T}$ stands for the transition matrix
\cite{reichl1998}, with time independent solution
\begin{equation}
\lvert \pi(t)\rangle  = \mathe^{-\hat{H} t} \lvert \pi(0)\rangle.
\label{eq:solution_markov}
\end{equation}

Despite the existence of this exact solution, the applicability of
Eq.~(\ref{eq:solution_markov}) at this stage is limited to small $N\sim
O(20)$. The reason is the exponential growth of the underlying vector
space as $2^N$. Here we show algorithms to generate the operators
$\hat{T}$ and $\hat{H}$ using finite symmetries or, equivalently,
permutation symmetries \emph{via} Cayley's theorem
\cite{hamermesh1962}. These algorithms are usually applied to
condensate matter physics \cite{artigo1,nakamuraPhysicaA2016} but due to
recent developments  
in the disease spreading dynamics \cite{nakamuraArxiv2016}, they may
also be employed in epidemiology studies. For pedagogical reasons, we
first show how to  
build the complete $2^N$ vector space and the corresponding transition 
matrix. Next, we explore cyclic permutations to construct the
cyclic vector space, in which $\hat{T}$ is broken down into $N$ smaller
blocks. Lastly, we consider the most symmetric cases, which reduce the
problem to O($N$). These instances correspond to the mean field or
averaged networks. 
The iteration of sparse $\hat{T}$ over $\lvert \pi(t)\rangle$ produces the
desired disease evolution among agents.
Relevant steps are shown in Algorithm \ref{algo1}. Numerical codes are
shown in pseudocode and Python.\footnote{Both Fortran and
  Python versions are available at
  https://github.com/gmnakamura/epidemic-transition-matrix.}

\section{Transition matrix}

The transition matrix $\hat{T}$ for SIS model \cite{nakamuraArxiv2016}
considering $N$ two-state agents is 
\begin{equation}
  \hat{T}=\mathbbm{1}- \beta\sum_{k j} \left[   A_{j k} ( 1
    -\hat{n}_j - \hat{\sigma}^{+}_j) + \Gamma \delta_{k
      j}(1-\hat{\sigma}^{-}_j) \right] \hat{n}_k,
\label{eq:transition_sis}
\end{equation}
where $\Gamma = \gamma  / \beta$, $\delta_{k l}$ is the Kronecker
delta,
\begin{equation}
  \hat{n}_k        \lvert  n_k  \rangle= n_k   \lvert  n_k  \rangle,
\end{equation}
is the localized number of infected operator ($n_k=0,1$) and
\begin{align}
  \hat{\sigma}_k^+ \lvert  n_k  \rangle=& \delta_{n_k,0} \lvert  1_k  \rangle,\\
  \hat{\sigma}_k^- \lvert  n_k  \rangle=& \delta_{n_k,1} \lvert  0_k  \rangle,
\end{align}
are Pauli raising and lowering localized operators, respectively. 
Local algebraic relationships are
$[\hat{n}_k,\hat{\sigma}^{\pm}_{k\prime}] = \pm\delta_{k\prime k}$ and
$[\hat{\sigma}_k^+,\hat{\sigma}_{k\prime}^-]= \delta_{k\prime
  k}(2\hat{n}_k-\mathbbm{1})$. 
Inspection of Eq.~(\ref{eq:transition_sis}) readily shows $\hat{T}$ is
not Hermitian. This means left- and right-eigenvectors are not related by
 Hermitian conjugation. In this scenario, the correct time
evolution of $\pi_{\mu}(t)$ using Eq.~(\ref{eq:solution_markov}) requires
the complete eigendecomposition, i.e. $2^N$ eigenvalues accompanied by
$2^N$ right-eigenvectors and $2^N$ left-eigenvectors. This is often the
main criticism against ABEM \cite{satorrasRevModPhys2015}.

\begin{figure}
\begin{tabular}{ccc}
\includegraphics[width=0.45\columnwidth]{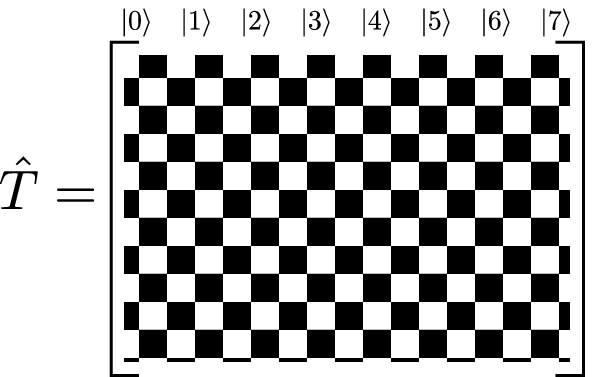} &\, &
\includegraphics[width=0.45\columnwidth]{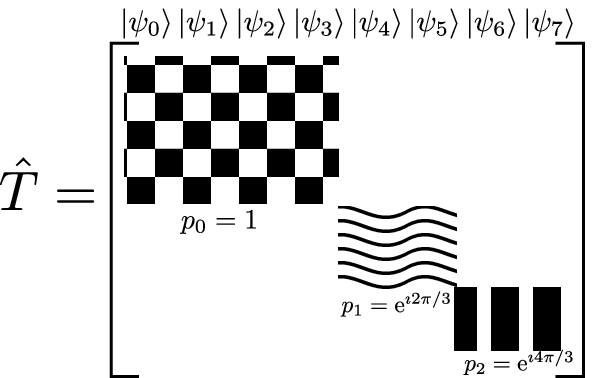}\\
a)& &b)
\end{tabular}
\caption{\label{fig_blocks}Reduction of transition matrix to block
  diagonal form. a) In the configurational vector space, $\{\lvert
  \mu\rangle\}$, the matrix representation of $\hat{T}$ lacks an explicit
  mathematical pattern. b) The emergence of organizational patterns are
  observed whenever symmetries of $\hat{T}$ are properly addressed by
  employing the eigenvectors $\{\psi\}$ and eigenvalues $\{\lambda\}$
  corresponding to the symmetry group considered. Under the invariant
  basis $\{\psi\}$, the matrix representation of $\hat{T}$ is brought
  to a block diagonal form, with blocks labeled by eigenvalues $\{\lambda\}$. 
} 
\end{figure}

However, the scenario described above is not entirely correct. The
rationale assumed all eigenstates are equally relevant, which is
incorrect whenever $A$ exhibits invariance upon the action of groups
(sets of transformations). Symmetries allow for the matrix representation
of $\hat{T}$ in block diagonal form, as shown in
Fig.~\ref{fig_blocks}. Eigenvectors related to each block share the same  
eigenvalue (degeneracy), as usual in quantum mechanics
\cite{sakurai1994}. Therefore, the trick lies in selecting the
appropriate base in respect to a given symmetry, redirecting
computational efforts towards smaller blocks, which is always more
efficient than working directly with the full matrix.

In the SIS model, cure events result from actions of one-body
operators, $\hat{\sigma}_k^- \hat{n}_k\equiv \hat{\sigma}_k^-$, on
configuration vectors. Infection events are 
two-body operators: one infected agent may transmit the communicable
disease to a susceptible agent after interaction between them, in the
time interval $\delta t$. Interestingly, the resulting interaction also
depends on symmetries available to the adjacency matrix $A$. The
symmetries available to $A$ may be further explored to assemble the
initial vector space, reducing $\hat{T}$ to its block diagonal form.

\begin{figure}
\begin{tabular}{ccc}
\includegraphics[width=0.40\columnwidth]{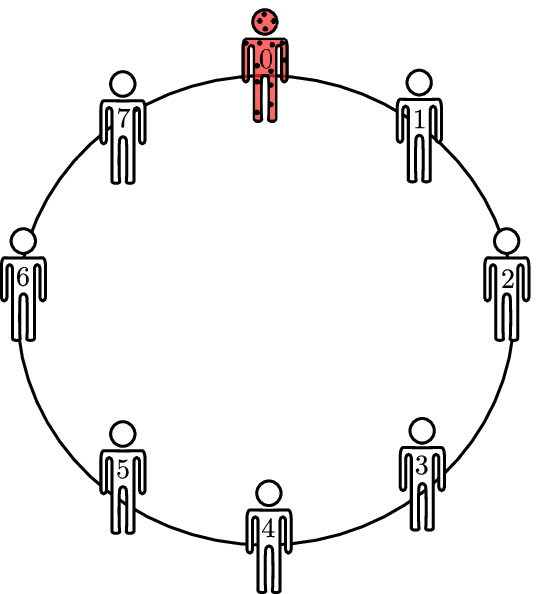}& $\quad$ &
\includegraphics[width=0.40\columnwidth]{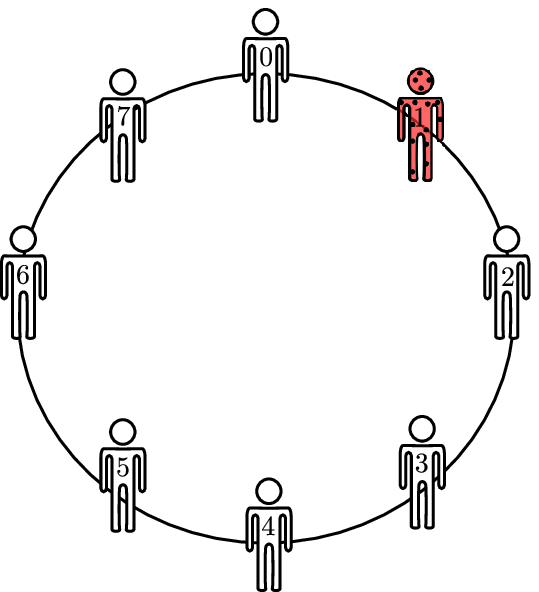}\\
a) & & b)
\end{tabular}
\caption{\label{fig_network_circular} Regular graph in circular
  representation for $N=8$ and single infected agent (red dotted). a) The
  infected 
  agent lies at node $k=0$. b) Graph obtained from cyclic permutation
  of nodes $k\rightarrow k+1$ and $N-1 \rightarrow 0$. Connections
  remain unchanged.} 
\end{figure}

Group operations over $A$ are always finite transformations. One may
explore the isomorphism between finite groups and the permutation
group \emph{via} Cayley theorem \cite{hamermesh1962} to build
permutation invariant subspaces. To that end, one must select the finite
group and the corresponding symmetry. For graphs, the circular
representation provides a convenient context to explore the existing
symmetries as Fig.~\ref{fig_network_circular}a) depicts. From
Fig.~\ref{fig_network_circular}b), connections among agents remain unchanged
after cyclic permutation of agents, hence, $A$ exhibits invariance under
cyclic permutations. 
Cyclic permutations form a subset of
permutation group and often represents geometric transformations such
as rotations and translations. 

Vectors with $N$ agents and invariant by cyclic permutations are built as
follows. Consider the \emph{representative} vector 
\begin{equation}
\lvert \mu_p\rangle \equiv 
\frac{1}{{\mathcal{N}_{\mu}}}\sum_{k=0}^{N-1}{\left( \mathe^{2 \imath \pi
    p/N}\hat{P} \right)}^k\lvert \mu\rangle,
\label{eq:representative}
\end{equation}
where $\mathcal{N}_{\mu}$ is the normalization. For clarity sake the integer
$p=0,1,\ldots,N-1$ is used to label the eigenvalue sector. The representative vector
$\lvert \mu_p \rangle$ describes the linear combination of
$N$-agent configurations related to $\lvert \mu\rangle$ by cyclic
permutations. For instance, $ \lvert 3_0 \rangle =\left( \lvert 011
  \rangle+\lvert 110 \rangle+\lvert 101 \rangle \right)/{\sqrt{3}}$
corresponds to the representative vector for $\mu=3$, with $N=3$ in the
$p=0$ sector.
By construction, the vectors $\lvert \mu_p\rangle$ satisfy the eigenvalue
equation
\begin{equation}
\hat{P}\lvert \mu_p\rangle = \mathe^{-2\imath \pi
  p/N}\lvert \mu_p\rangle.
\label{eq:eigenequation_cyclic}
\end{equation}
The eigenvalues $\mathe^{-2\imath  \pi p/N}$ are derived from
$\hat{P}^{N}=\mathbbm{1}$. Since cyclic permutations never change link
distributions, only node labels, cyclic permutation eigenvectors are
suitable candidates to reduce $\hat{T}$ to block diagonal form whenever
$[\hat{P},\hat{T}]=0$.



\section{Cyclic vector space}

The complete picture of infection dynamics generated by SIS model
requires the utilization of $2^N$ configuration vectors. For completeness
sake, we discuss the algorithm to obtain the vector space using both
string and numeric representations. Matrix elements of $\hat{T}$ in
Eq.~(\ref{eq:transition_sis}) are calculated from  adjacency matrix and
user input dictionary (lookup table) based on off-diagonal transition rules.

\begin{figure}
  \includegraphics[width=0.5\columnwidth]{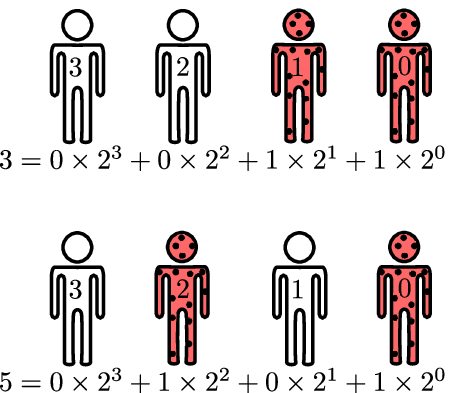}
  \caption{\label{fig_conf} Agent configurations using binary
    representation for $\mu=3$ and $5$ with $N=4$. For $\lvert \mu=3
    \rangle = \lvert 0011 \rangle$ whereas $\lvert \mu=5
    \rangle = \lvert 0101 \rangle$. In both configurations,  two
    agents are infected (red dotted). }
\end{figure}

According to Eq.~(\ref{eq:integer_representation}), the configuration
vector $\lvert \mu \rangle$ is obtained from the binary
representations of the labels $\mu$, as exemplified in
Fig.~\ref{fig_conf}. There are two common equivalent routes
to implement the configuration in computer codes. The first method employs
string objects whereas the second method makes use of
discrete mathematics. The second approach tends to be more
efficient for two-state problems as optimized and native libraries for
binary operations are widely available.\footnote{For two-state
  variables, binary logical operations and binary manipulation inherently
  produce pipeline parallelism.} For pedagogical purposes and
generalization for more than two-states, we avoid exclusive binary
operations in favor of usual discrete integer division and modulo operations. 

In Python, classes provide a convenient mechanism to enable both
formalisms for each instanced object (vector). Here, the custom class
SymConf is used to encapsulate two instance variables: \emph{label}
stores the string representation of $N$ agents, while \emph{label\_int}
stores the corresponding integer number. In addition, the custom class also
encapsulates three global class variables, \emph{base, dimension} and
\emph{basemax} whose default values are $2$, $N$ and $2^N$. Base
corresponds to the number of available states per agent. The class
main method generates the eigenvectors $\lvert \mu_p\rangle$
with eigenvalue $\exp(-2\imath\pi p/N)$, relative to the cyclic permutation
operator $\hat{P}$ using Eq.~(\ref{eq:representative}). 

In what follows, we address four relevant points regarding the permutation
eigenvectors $\lvert \mu_p\rangle$, namely, the criteria used to label
eigenvectors; normalization; number of infected agents; and the
permutation operation.


\emph{Labels}. Eq.~(\ref{eq:representative}) claims permutation
eigenvectors are linear combination of all configuration vectors related
by cyclic permutations. Here we set the 
{convention} to adopt the smallest value $\mu$ present in the linear
combination to label the representative vector. As examples, consider
the following representatives of $\mu=1$, $N=4$ and $p=0,1,2,3$:
\begin{align}
\lvert 1_0 \rangle=& \frac{\lvert 0 0 0 1 = 1\rangle+\lvert 0 0 1 0 =2\rangle+\lvert 0
1 0 0 = 4\rangle+\lvert 1 0 0 0 = 8\rangle}{\sqrt{4}}.\label{eq:state1}\\
\lvert 1_1 \rangle=& \frac{\lvert 1\rangle+\imath \lvert 2\rangle-\lvert
                     4\rangle-\imath\lvert 8\rangle}{\sqrt{4}}.\\
\lvert 1_2 \rangle=& \frac{\lvert 1\rangle - \lvert 2\rangle+\lvert
                     4\rangle - \lvert 8\rangle}{\sqrt{4}}.\\
\lvert 1_3 \rangle=& \frac{\lvert 1\rangle -\imath \lvert 2\rangle-\lvert 4\rangle+\imath\lvert 8\rangle}{\sqrt{4}}.
\end{align}
The order convention is necessary to calculate the relative phase
between configurations related by permutations, in non-trivial linear 
combinations. Consider 
\begin{equation}
\lvert \phi\rangle =\hat{P}\lvert\mu_p\rangle=
\frac{1}{\mathcal{N}_{\mu}}\sum_k(\mathe^{2\imath \pi p   /N}\hat{P})^k
\hat{P}\lvert\mu\rangle.
\end{equation}
Since $\lvert \mu\rangle$ and
$\hat{P}\lvert\mu\rangle$ are related by a single cyclic permutation,
$\lvert \phi\rangle = \mathe^{-2\imath\pi p /N} \lvert \mu_p
\rangle$. Note that the linear combination $\hat{P}\lvert {\mu_p}\rangle+\lvert
{\mu_p}\rangle= (1+\mathe^{-2\imath\pi   p/N})\lvert {\mu_p}\rangle $
vanishes for $p=N/2$.  Despite the simplicity of the previous example,
it already illustrates the relevance of phase difference among cyclic vectors.

\emph{Normalization}. According to Eq.~(\ref{eq:representative}), the
squared norm of representative vectors is
\begin{equation}
\langle \mu_p\vert \mu_p \rangle =
\frac{1}{\mathcal{N}_{\mu}}\sum_{k=0}^{N-1} \mathe^{-2\imath \pi
  p k/N}\langle \mu \rvert \hat{P}^{-k}  \lvert {\mu_p}\rangle=
\frac{N}{\mathcal{N}_{\mu}} \langle \mu \vert \mu_p\rangle. 
\label{eq:scalar_product}
\end{equation}
The evaluation of the scalar product $\langle \mu \vert \mu_p\rangle$
follows directly from Eq.~(\ref{eq:representative}). 
One notices the configuration $\lvert \mu\rangle$ may appear only
once for several linear combinations $\lvert \mu_p\rangle$, so that $\langle
\mu \vert \mu_p\rangle=1/\mathcal{N}_{\mu}$. For instance, this is the case of
$\langle 1\vert 1_p\rangle$. However, a given configuration
$\lvert\mu\rangle$ may contribute more than once if there exist an
integer $1\le r \le N$ such that $\hat{P} ^{r}\lvert \mu\rangle = \lvert
\mu\rangle$, i.e., after $r$ cyclic permutations the configuration
repeats itself. Since  $\hat{P}^N=\mathbbm{1}$, it follows $N/r$ is the
number of times the configuration $\lvert \mu\rangle$ appears in $\lvert
\mu_p\rangle$. Each contribution adds $\mathe^{2\imath \pi
p m/N }/\mathcal{N}_{\mu}\, (m=0,1,\ldots,N/r-1)$  in 
Eq.~(\ref{eq:scalar_product}). This result is conveniently summarized
using the repetition number
\begin{equation}
R_{\mu,p} =  \sideset{}{'}\sum_{m=0}^{N/r-1}(\mathe^{2\imath \pi p \,r/N
})^m,
\label{eq:r_mu}
\end{equation}
where the primed sum indicates $N/r$ in the upper limit is an integer
number. Therefore, $\langle \mu \vert  \mu_p\rangle =
R_{\mu,p}/\mathcal{N}_{\mu}$ and one obtains 
\begin{equation}
\mathcal{N}_{\mu}= \sqrt{N  R_{\mu,p}}
\end{equation}
 from Eq.~(\ref{eq:scalar_product}). 

We now show two examples to consolidate the discussion around
$R_{\mu,p}$ and $\mathcal{N}_{\mu}$, for $N=4$ and two infected
agents. The configuration state 
$\lvert 3 \rangle=\lvert 0011\rangle$ requires $N$ cyclic permutations
to repeat itself, so 
that $R_{3 , p}=1$ for any $p$ and the corresponding normalization for
$\lvert 3_p\rangle$ is simply $\mathcal{N}_{3}=\sqrt{N}$, as
expected. The first non-trivial case arises for $\lvert 
5_p\rangle$ because the base configuration $\lvert 5 \rangle = \lvert
0101\rangle$ satisfies $\hat{P}^2\lvert 5\rangle = \lvert
5\rangle$. According to Eq.~(\ref{eq:r_mu}),
$R_{5,p}=1+\mathe^{4\imath\pi p /N}$  and assume only values:
$R_{5,0}=R_{5,2}=2$ and $R_{5,1}=R_{5,3}=0$. 
 Thus, depending on $p$,  linear combinations are \emph{forbidden}
(null-normed vectors), ensuring the correct dimension of vector
space. The remaining non-null states for $N=4$ are shown in
Table~\ref{table:eigenvectors} for further reference.

\begin{table}
\caption{\label{table:eigenvectors} Cyclic permutation eigenvectors with
  $N=4$ agents. The first column shows the number of infected agents in
  the eigenvector. Each remaining column corresponds to a permutation
  sector $p$, and each row the corresponding state $\lvert
  \mu_p\rangle$. The cross symbol indicates null-normed vector and the
  dimension of the vector space is $d=2^4$.} 
\begin{ruledtabular}
\begin{tabular}{lllll}
$n$& $p=0$ & $p=1$ & $p=2$ & $p=3$ \\
\hline
 $0$
& $\lvert 0_0 \rangle$ 
        &  $\times$
                & $\times$
                        & $\times$
  \\
  $1$ 
& $\lvert 1_0 \rangle$
        & $\lvert 1_1  \rangle$
                &$\lvert 1_2  \rangle$ 
                        &  $\lvert 1_3 \rangle$
  \\
  $2$  
&$\lvert 3_0 \rangle$
        & $\lvert 3_1 \rangle$
                &$\lvert 3_2 \rangle$
                        & $\lvert 3_3 \rangle$
  \\ 
  $2$ 
& $\lvert 5_0 \rangle$
        & $\times$
                & $\lvert 5_2 \rangle$ 
                        & $\times$\\
  $3$ 
&  $\lvert 7_0 \rangle$
        &  $\lvert 7_1 \rangle$
                & $\lvert 7_2 \rangle$
                        & $\lvert 7_3 \rangle$
  \\ 
$4$ 
& $\lvert 15_0 \rangle$ 
        & $\times$
                & $\times$
                        &$\times$
\end{tabular}
\end{ruledtabular}
\end{table}

\emph{Number of infected agents}.  The number of infected agents using
 representative vectors is
\begin{equation}
\langle \hat{n} \rangle_{\mu} = \sum_k \langle \mu_p \rvert
\hat{n}_k\lvert \mu_p\rangle.
\end{equation} 
In the string representation, native string methods such as
\emph{count('x')} count the number agents with health state
$x=0,1,2\ldots$. 
If native methods are unavailable, one may always perform
a comparative loop over the string. Algorithm~\ref{algo_count} explains
the standard procedure to count bits in the integer representation.

\begin{figure}
\begin{tabular}{ccc}
\includegraphics[width=0.45\columnwidth]{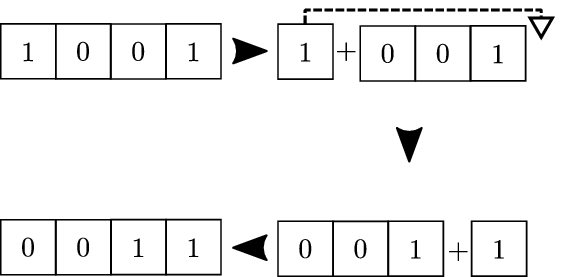} & $\quad$ &
\includegraphics[width=0.45\columnwidth]{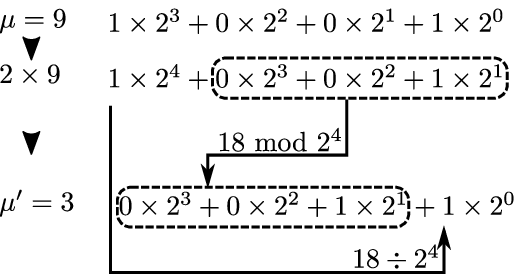} \\
a) & & b)
\end{tabular}
\caption{\label{fig_permutation} Cyclic permutation for configuration
  $\mu=9$ with $N=4$. a) in string representation executes one copy
  and one concatenation operation; b) integer representation requires
both integer division and modulo operation by $2^N$.}
\end{figure}

\emph{Permutation}. Cyclic permutations are the core transformations
here. In the string representation, cyclic permutations consist 
of one copy and one concatenation call, 
as exemplified in Fig.~\ref{fig_permutation}a).
Meanwhile, in the integer representation, cyclic permutations are
obtained using modulo and integer division: 
$\mu^{\prime} = (2 \mu$ \% $2^N) + (2\mu // 2^{N})$,
the new configuration $\mu^{\prime}$ is obtained from configuration
$\mu$ taking the 
modulo of $2 \mu$ by $2^N$ in addition to the result of the integer
division $2 \mu$ by $ 2^{N} $. Multiplication by the number of available
states translates bit fields to the left. The modulo operation crops
contributions larger than those available to $N$-bit fields. Integer
division  $2 \mu/2^{N}$ selects the bit associated to largest binary
position and shifts it to the lowest binary position. See
Fig.~\ref{fig_permutation}b).

Next, we focus attention only to sector $p=0$, as it holds both 
all-infected and all-cured representative vectors. This route allows
for the exact evaluation of $\pi_{2^N-1}$, the probability the disease has
reached every agent in the system; or the evaluation of $\pi_0$, the
disease eradication probability. Roughly speaking, the $p=0$
sector also holds the largest dimensionality. 
Consider each integer $\mu$ in $[0,2^N)$ as a potential candidate to
assemble the symmetric vector spaces for fixed $p$. By performing
$N-1$ cyclic permutations over $\lvert\mu\rangle$, one determines the
representative state $\lvert\mu_p\rangle$ in
Eq.~(\ref{eq:representative})  as well as the number of
repetitions $R_{\mu,p}$, hence the norm
$\mathcal{N}_{\mu}$. Algorithm~\ref{algo_representative} calculates the
representative vector $\lvert \mu_p\rangle$ associated with
configuration $\lvert \mu\rangle$.  
Due to the order convention adopted here, the string representation
must be converted to the integer representation at the
\emph{if}-clause test.
The representative configurations are then stored either in a list or
dictionary. As additional benefit, since vector spaces are independent on
the problem at hand, the set of representatives may also be stored in a
database for further use in different problems, as long as they are
subjected to the same symmetry.

\section{Matrix elements}

The next step is the evaluation of $\hat{T}$ in the $p=0$
sector. Infection and cure dynamics are the main actors in this context
as they inform the way representative vectors $\lvert \mu_0 \rangle $
interact with each other, $\hat{T}\lvert \mu_0 \rangle =
\sum_{\{\nu\}}^{\prime}\,T_{\nu\mu}\lvert \nu_0 \rangle$. The prime
indicates the sum runs over all eigenvectors in the $p=0$ sector, while
cyclic permutation invariance implies   
\begin{equation}
\hat{T}\lvert \mu_0\rangle =
\frac{1}{\mathcal{N}_{\mu}}\sum_{k=0}^{N-1}\hat{P}^k\,\hat{T}\lvert
\mu\rangle.
\label{eq:tpsi}
\end{equation}
Eq.~(\ref{eq:tpsi}) tell us the action of $\hat{T}$ on the linear
combination $\lvert \mu_0\rangle$ is calculated from the
simpler operation $\hat{T}\lvert \mu\rangle$. The resulting vectors
are then permuted, producing the corresponding matrix elements. For
instance, consider $\hat{T}\lvert 7_0 \rangle$ for $N=3$:
\begin{align}
\hat{T}\lvert 7_0\rangle
  =&\frac{1}{\mathcal{N}_7}\sum_{k=0}^{2}\hat{P}^k
    \hat{T}\lvert 7\rangle 
  =\frac{\gamma}{\mathcal{N}_7}\sum_{k=0}^{2}\hat{P}^k\left(
    \lvert 3\rangle+\lvert 5\rangle+\lvert 6\rangle \right)\nonumber\\
  =&\frac{\gamma}{\mathcal{N}_7}\sum_{k=0}^{2}\hat{P}^k\left(
     \lvert{3}\rangle+\hat{P}\lvert{3}\rangle+\hat{P}^2\lvert 3\rangle
\right)=
\left(3\gamma\frac{\mathcal{N}_{3}}{\mathcal{N}_{7}}\right) \lvert
    3_0\rangle\nonumber\\
  =&\sqrt{3}\gamma \lvert 3_0\rangle.
\end{align}

\begin{figure}
\includegraphics[width=0.5\columnwidth]{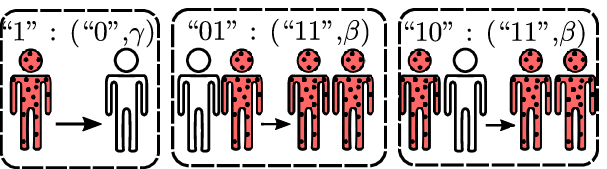}
\caption{\label{fig_rules}Off-diagonal transitions in the SIS
  model. Data structure follows the income-outcome convention. Data
  entries represent the current one-body (two-body) health state whereas the
  corresponding data values, organized as tuples, express the outcome one-body
  (two-body) configuration and coupling strength.}
\end{figure}

The relevant data structure for $\hat{T}$ are the off-diagonal
transitions, which are further subdivided into two categories: one or
two-body contributions. This is illustrated in Fig.~\ref{fig_rules} for
the SIS model. The finite set of transition rules are passed
as a lookup table or, if available, a dictionary. Data is
organized as follows: each entry represents one or two-body
configuration whose value corresponds to one tuple. Each tuple holds
two immutable values: the configuration to which the entry transitions
to and the assigned coupling strength. 

\begin{figure}
\includegraphics[width=0.45\columnwidth]{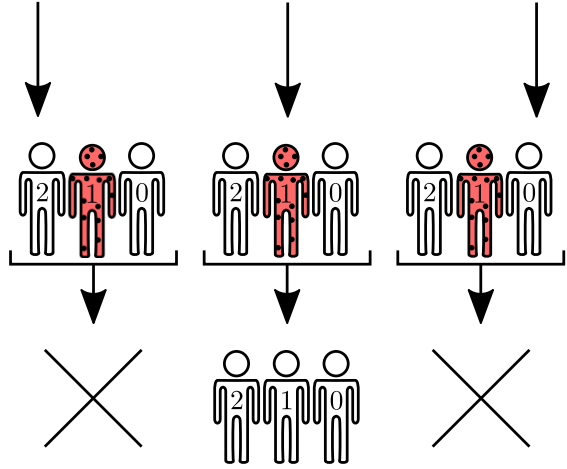}
\caption{\label{fig_onebody}Cure operator action on configuration vector
  $\lvert 2\rangle$, in the SIS model with $N=3$ . Non-vanishing
  transition is observed only for agent $k=1$, which is infected,
  producing $\lvert 0 \rangle$.} 
\end{figure}

With off-diagonal transition rules in hand, one-body actions are
evaluated by scanning each agent and applying the corresponding
transition rule in Algorithm~\ref{algo_onebody}. The resulting one-body
transitions are stored in the \emph{outcome}
variable. Fig.~\ref{fig_onebody} depicts an example for $N=3$ and one
infected agent at $k=1$.

\begin{figure}
\includegraphics[width=0.45\columnwidth]{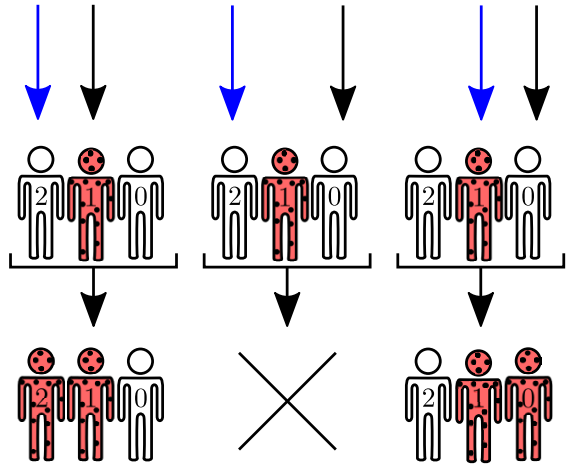}
\caption{\label{fig_twobody}Infection operator action on configuration
  vector $\lvert 2\rangle$, in the SIS model with $N=3$ and mean field
  network. Disease transmission events are evaluated for each pair of
  agents. Whenever the pair health state differs, and the pair also
  shares one connection expressed by the adjacency matrix, the
  configuration changes to contemplate the recently infected
  individual. For $\lvert 2\rangle$, $k=1$ agent contaminates $k=0$
  ($k=2$) agent, producing the configuration $\lvert 3\rangle$ ( $\lvert
  6\rangle$ ).}   
\end{figure}

Two-body operators differ from their one-body counterparts due to the fact they
require two agent loops and information from the adjacency matrix $A$, as seen in the
Algorithm~\ref{algo_twobody}. Fig.~\ref{fig_twobody}
exhibits an example for $N=3$. After both one- and two-body transitions
are computed, the diagonal element is obtained \emph{via} probability
conservation: $T_{\mu\mu}=1-\sum_{\mu\neq \nu}^{\prime}T_{\mu\nu}$. The
process is iterated until all eigenvectors and their respective
transitions are accounted for.

\section{Casimir vector space}

The recent advances in the disease spreading dynamics in realistic
populations are intimately linked to network theory
\cite{satorrasRevModPhys2015}. In this context, a network corresponds to
an ensemble of graphs sharing common characteristics, whose adjacency
matrix occurs according to the network probability distribution function
(NPDF). In this sense, a graph is one sample or realization of the
network. Statistical properties of networks are derived for each graph
and then taking the ensemble average and 
deviation. In practice, when graphs in the ensemble are large enough ($N\gg
1$) and representatives, statistics may also be evaluated for each graph
and extrapolated as those of the network. 

Two cases hold particular importance for applications of network theory in
epidemic models: the mean field and random networks. In the first case,
all agents are connected to each other, meaning one infected agent may
potentially infect anyone. Hence, the disease tends to spread faster
than in constrained networks. Furthermore, all graphs in the mean field
ensemble share the same adjacency $A^{\text{MF}}$. In the other case,
the connection between agent $i$ and $j$ occurs with fixed probability
$\rho$. However, graphs in the random network ensemble differ from each
other. Here, we only consider ensemble averages as a way to extract
statistical properties, which is equivalent to set $A_{i
  j}^{\text{random}}=\rho\,(1-\delta_{ij})=\rho A_{i
  j}^{\text{MF}}$. Thus, all relevant symmetries lie only in the mean
field adjacency matrix $A^{\text{MF}}$.
Naturally, $A^{\text{MF}}$ remains invariant under cyclic
permutations, enabling the application of the algorithm explained in the
previous sections. However, $A^{\text{MF}}$ is also symmetric under the
action of any permutation, which drastically reduces the diagonal blocks
 of $\hat{T}$ from $O(2^N/N)$ to $O(N)$.

Here, our main concern is to employ the cyclic permutation eigenvectors
$\lvert \mu_p\rangle$ to generate the eigenvectors of the complete
permutation group, $\lvert s, m; p\rangle$. The eigenvectors $\lvert
s, m; p\rangle$ reduce $\hat{T}$ in mean field or random networks to
block diagonal form with dimension $O(N)$. The indices $s$ and $m$ may
assume the following values $s=N/2,N/2 - 1,\ldots$ with $s>0$ and
$m=-s,-s+1,\ldots,s$, respectively. Clearly, the relationship between
$s$ and $m$ are the same as those observed for quantum spin
operators. The explanation goes as follows. As shown in Ref.~\cite{nakamuraArxiv2016},
Eq.~(\ref{eq:transition_sis}) in either mean field or random networks 
contains operators $\hat{S}^{\pm}\equiv \sum_k\hat{\sigma}^{\pm}_k$ and
$\hat{n} \equiv \sum_{k}\hat{n}_k$. From the important relation
$\hat{n}=\hat{S}^z+N/2$, one retains spin operators and the upper
bound $s=N/2$, as expected from the combination of $N$ $1/2$-spin
particles.

In what follows, we only consider the $p=0$ sector. First, let
$\hat{S}^2=(\hat{S}^{z})^2+(\hat{S}^{+}\hat{S}^{-}+\hat{S}^{-}\hat{S}^{+})/2$
be the Casimir operator, so that $[\hat{S}^{2},\hat{S}^{\alpha}]=[\hat{P},\hat{S}^{\alpha}]=0$ for
$\alpha=z,\pm$ and $\hat{S}^{2}\lvert s, m; 0 \rangle = s(s+1)\lvert s,
m; 0 \rangle$. Accordingly, $[\hat{S}^2,\hat{T}]=0$ and $s$ and $p$ are good
quantum numbers. In general, the eigenvector $\lvert s,m;p\rangle$ may
always be expressed as
\begin{equation}
\lvert s, m; p \rangle = \sum_{\mu} c_{\mu}^{smp} \lvert
\mu \rangle. 
\label{eq:tmp2}
\end{equation}
Clearly, $c_{\mu}^{smp}=0$ if the number of infected agents in the
configuration $\mu$, $n_{\mu} = \sum_k \langle \mu_0 \vert \hat{n}_k
\vert \mu_0 \rangle $, fails to satisfy the constraint $n_{\mu} =
m+N/2$. The idea is to write Eq.~(\ref{eq:tmp2}) as a linear combination
of representative vectors $\lvert \mu_p \rangle$ with $m+N/2$ infected
agents, ensuring all available permutations are accounted for. The
implications for numerical codes is quite obvious: it allows the
reutilization of numerical codes to obtain eigenvectors $\lvert \mu_p
\rangle$.

The most relevant sector for epidemic models contains the configuration
with all (none) infected agents. According to previous sections, this
implies $p=0$ while $m=\pm N/2$ requires $s=N/2$. In the $(s=N/2,p=0)$
sector, the desired linear combination is   
\begin{equation}
\lvert s=N/2,m,p=0\rangle =
\frac{1}{\mathcal{N}}\sideset{}{'}\sum_{\{\mu\}}
{R_{\mu,0}}^{-1/2}{\lvert \mu_0 \rangle},
\label{eq:linear_combination}
\end{equation}
with normalization $\lvert\mathcal{N}\rvert ^2=\sum^{\prime}_{\mu}\lvert
R_{\mu,0}\rvert ^{-1}$. The
prime indicates the sum is subjected to the constraint $n_{\mu}=m+N/2$
for $m=-N/2,\ldots,N/2$. The result in Eq.~(\ref{eq:linear_combination})
agrees with the standard theory of spin addition. Generalization for
 $p$ and $s$ is straightforward and omitted. It is worth
mentioning the formalism adopted here already accounted for forbidden
states in $p\neq 0$ sectors.

Examples are available to appreciate Eq.~(\ref{eq:linear_combination})
for increasing values of $N$. We begin by considering $N=4$. This
translates into $s=2$ and $m=-2,\ldots,2$. The relevant representative
eigenvectors $\lvert \mu_0\rangle$ are expressed in
Table~\ref{table_eigenvectors1}.  The only non-trivial correspondence
occurs for $m=0$,
\begin{align}
\lvert 2,0; 0 \rangle =& \frac{\sqrt{2}\lvert 3_0 \rangle +\lvert
                         5_0\rangle}{\sqrt{3}},\\
=&\frac{\lvert 0011\rangle+\lvert 1001\rangle+\lvert 1100\rangle+\lvert 0110\rangle+\lvert 0101\rangle+\lvert 1010\rangle}{\sqrt{6}}.\nonumber
\end{align}
Next, consider $N=6$ which fixes $s=3$ and $m=-3,\ldots,3$. The
eigenvector $\lvert 3 , 0; 0\rangle$ holds contributions from four
cyclic eigenvectors or, equivalently, $20$ configurations:
\begin{align}
\lvert 3,0; 0 \rangle =& \frac{\sqrt{3}\lvert 7_0 \rangle+
                         \sqrt{3}\lvert 11_0 \rangle +
                         \sqrt{3}\lvert 19_0\rangle +
                         \lvert 21_0\rangle}{\sqrt{10}},\\
=&\frac{
   \lvert 000111\rangle+\lvert 100011\rangle+\lvert 110001\rangle+
   \lvert 111000\rangle+\lvert 011100\rangle+\lvert 001110\rangle
}{\sqrt{20}}.\nonumber\\
+&\frac{
   \lvert 001011\rangle+\lvert 100101\rangle+\lvert 110010\rangle+
   \lvert 011001\rangle+\lvert 101100\rangle+\lvert 010110\rangle
}{\sqrt{20}}.\nonumber\\
+&\frac{
   \lvert 010011\rangle+\lvert 101001\rangle+\lvert 110100\rangle+
   \lvert 011010\rangle+\lvert 001101\rangle+\lvert 100110\rangle
}{\sqrt{20}}.\nonumber\\
+&\frac{
   \lvert 010101\rangle+\lvert 101010\rangle
}{\sqrt{20}}.\nonumber
\end{align}

\begin{table}
\caption{\label{table_eigenvectors1} Eigenvectors $\lvert \mu_0\rangle$ with $N=4$.}
\begin{ruledtabular}
\begin{tabular}{c c r c}
$\mu_0$&$R_{\mu,0}$&$m$&$\lvert{\mu}\rangle$\\
\hline
 $0$&$4$&$-2$&$\lvert 0 0 0 0\rangle$\\
 $1$&$1$&$-1$&$\lvert 0 0 0 1\rangle$\\
 $3$&$1$&$0$&$\lvert 0 0 1 1\rangle$\\
$5$ &$2$&$0$&$\lvert 0 1 0 1\rangle$\\
$7$ &$1$&$1$&$\lvert 0 1 1 1\rangle$\\
$15$&$4$&$2$&$\lvert 1 1 1 1\rangle$
\end{tabular}
\end{ruledtabular}
\end{table}

\section{Discussion}

The algorithms presented in this study assumed only two health states
for each agent. Generalization for $q$ number of states is
readily available by changing to the integer representation,
\begin{equation}
\mu = a_{N-1} q^{N-1}+\cdots+a_0 q^0,
\end{equation}
with $a_k=0,1,\ldots,q-1$, concomitant with additional off-diagonal
transitions. For instance, the SIRS ABEM generalizes the SIS model as it
introduces the removed (R) health state for agents. This additional
state often means the agent has recovered from the illness and developed
immunity, has been vaccinated or has passed away. In any case, once
removed, the agent takes no 
part in the dynamics of disease transmission, hindering infection events
\cite{satorrasRevModPhys2015}. As such, cure with immunization or death events produce
the transition $I\rightarrow R$, with probability $\gamma$ while vaccination
$S\rightarrow R$ occurs with probability $\xi$. To reflect the updated
one-body off-diagonal transition, the following transition rules
dictionary is used:
\begin{Verbatim}[commandchars=\\\{\},codes={\catcode`\$=3\catcode`\^=7\catcode`\_=8}]
\PYG{n}{rules}\PYG{o}{=}\PYG{p}{\PYGZob{}}\PYG{l+s+s1}{\PYGZsq{}0\PYGZsq{}}\PYG{p}{:(}\PYG{l+s+s1}{\PYGZsq{}2\PYGZsq{}}\PYG{p}{,}\PYG{n}{xi}\PYG{p}{),}   \PYG{c+c1}{\PYGZsh{}$S \rightarrow R$}
       \PYG{l+s+s1}{\PYGZsq{}1\PYGZsq{}}\PYG{p}{:(}\PYG{l+s+s1}{\PYGZsq{}2\PYGZsq{}}\PYG{p}{,}\PYG{n}{gamma}\PYG{p}{),}\PYG{c+c1}{\PYGZsh{}$I \rightarrow R$}
      \PYG{l+s+s1}{\PYGZsq{}01\PYGZsq{}}\PYG{p}{:(}\PYG{l+s+s1}{\PYGZsq{}11\PYGZsq{}}\PYG{p}{,}\PYG{n}{beta}\PYG{p}{),}\PYG{c+c1}{\PYGZsh{}$SI\rightarrow II$}
      \PYG{l+s+s1}{\PYGZsq{}10\PYGZsq{}}\PYG{p}{:(}\PYG{l+s+s1}{\PYGZsq{}11\PYGZsq{}}\PYG{p}{,}\PYG{n}{beta}\PYG{p}{)\PYGZcb{}}\PYG{c+c1}{\PYGZsh{}$IS\rightarrow II$}
\end{Verbatim}
where $2$ represents state $R$. If death events are excluded,
temporary immunization is achieved via $R\rightarrow S$ with probability
$\eta$.

Parallelism merits further discussion. The computation of representative
vector space may be performed in parallel by dividing the set of  $q^N$
integers among $Q$ processes. Each process runs one local set of
representative vectors which, posteriorly, is compared against the sets from the
remaining processes. The union of all $Q$ sets produces the desired 
representative vector space. Parallelism is also obtained at the
evaluation of $\hat{T}$: columns ( $\lvert \mu_p\rangle $ ) are
distributed among $Q$ processes and the corresponding matrix elements are
calculated for each process. The union of all matrix elements from each
process produces the complete description of $\hat{T}$ in the
representative vector space. Lastly, parallelism is also available for
sparse products $\hat{T}\lvert \pi(t)\rangle $ necessary to execute the
time evolution.

We also emphasize the algorithms explained here are most useful to
evaluate quantities within a single permutation sector of
$\hat{T}$. This is likely the case whenever the probability for disease
eradication or complete population contamination are concerned. 
Another relevant situation occurs when the initial condition itself
falls within a single sector. For instance, the initial probability vector
$\lvert \pi(0)\rangle = (1/3)(\lvert 001\rangle + \lvert
010\rangle+\lvert 100\rangle)$ states only one among $N=3$ agents is
infected. However, the identity of the infected agent is unknown \emph{a
  priori}, so that configurations with one infected agent occurs
with equal probability $1/N$. Now, the decomposition of $\lvert
\pi(0)\rangle$ in the $\lvert \mu_p\rangle$ basis results in $\lvert
\pi(0)\rangle = (1/\sqrt{3})\lvert 1_0\rangle$. Thus, the time evolution
of $\lvert\pi(0)\rangle$ by the action of $\hat{T}$ is again restricted
to a single permutation sector. 

Without loss of generality, the initial condition can always be written as  
\begin{equation}
\lvert \pi(0)\rangle = \sideset{}{'}\sum_{\{\mu\}}\sum_{k=0}^{N-1}
{\pi}_{\mu k} \hat{P}^k \lvert \mu\rangle,
\end{equation}
where the primed sum runs only over the indices $\mu$, which
also labels the representative vectors. The cyclic permutation
$\hat{P}^k$ generates the remaining configurations related to $\lvert
\mu\rangle$ whereas the coefficients ${\pi}_{\mu k}$ are the
corresponding initial probabilities. From the eigenvalue equation 
Eq.~(\ref{eq:eigenequation_cyclic}), one calculates the scalar product
\begin{align}
\langle \nu_p\vert \pi(0)\rangle
  &=\sideset{}{'}\sum_{\{\mu\}}\sum_{k=0}^{N-1}{\pi}_{\mu k}\langle \nu_p \vert
    \hat{P}^k\vert \mu\rangle
  =\sideset{}{'}\sum_{\{\mu\}}\sum_{k=0}^{N-1}{\pi}_{\mu k}
    \mathe^{2\imath \pi p k /N}\langle \nu_p\vert 
    \mu\rangle\nonumber\\
  &=\sqrt{N} \tilde{\pi}_{\nu p} R_{\nu p}/\mathcal{N}_{\mu},
\end{align}
where $\tilde{\pi}_{\mu p} =N^{-1/2} \sum_k  {\pi}_{\mu k}
\mathe^{2\imath\pi p k /N}$ is the discrete Fourier transform of
${\pi}_{\mu k}$.
Using the previous example, with
one infected among $N=3$ agents, 
\begin{equation}
\lvert \pi(0)\rangle =  \sum_{k=0}^{2}{\pi}_{0 k} \hat{P}^k\lvert
0\rangle+
\sum_{k=0}^{2}{\pi}_{1 k} \hat{P}^k\lvert 1\rangle+
\sum_{k=0}^{2}{\pi}_{3 k} \hat{P}^k\lvert 3\rangle+
\sum_{k=0}^{2}{\pi}_{7 k} \hat{P}^k\lvert 7\rangle,
\end{equation}
with ${\pi}_{\mu k}= \delta_{\mu 1}/3$ so that $R_{1 p} = 1$,
$\tilde{\pi}_{1 p} = \delta_{p 0}/\sqrt{3}$, and the previous result is
recovered. 

We now address the case where the evaluation of the
desired statistics requires several permutation sectors. In the worst
case scenario, every permutation sector contributes equally to the
computation. Therefore one must diagonalize each block in order to
obtain the relevant eigenvalues and eigenvectors. As a crude
approximation, one may consider the $N$ blocks have the same dimension
$d/N$ for a $d$-dimensional vector space. The complexity of
 diagonalization methods in the LAPACK library range from $O((d/N)^2)$ up to
 $O((d/N)^3)$ for each block \cite{demmel2008}, whereas the complexity
 range for full  diagonalization is $[O(d^2),O(d^3)]$. Thus
 diagonalization of $N$ blocks reduces the total complexity from
 $N^{-1}$ up to $N^{-2}$. More importantly, because blocks are
 disjointed, they can be diagonalized in different processors.

As the closing remark, the algorithms presented here are most suitable for
networks with invariance by cyclic permutations. However, they are also
convenient whenever the algebraic commutator can be approximated by
$[\hat{T} , \hat{P}]=\hat{O}$, where the operator $\hat{O}$ is
symmetric under cyclic permutations, $[\hat{O},\hat{P}]=0$.  In
particular, $\hat{O}=q_{0} \mathbbm{1}+q_{1}
\hat{P}^{y}+\sum_{\beta=z,\pm}q_{\beta}\hat{S}^{\beta}$,  with constant
$q_{j}$ ($j=0,1,z,\pm$) and $y\in\mathbbm{R}$, creates interesting
disease spreading dynamics such as localized disease source for
$q_{\beta}=q\delta_{\beta,0}$.

\section{Conclusion}

Agent-based epidemic models describe disease spreading dynamics
in networks. Direct investigation of epidemic Markov processes is often
hindered due to the exponential increase of vector space dimension with the
number of agents. By exploiting cyclic permutation symmetries, relevant
elements to the dynamics are confined to a single permutation sector,
greatly reducing computation efforts. The $p=0$ sector holds particular
importance as it contains configurations where none or all agents are
infected. In practice, by selecting one cyclic permutation eigensector,
one extract relevant information rather than all available information. 
Moreover, cyclic permutation eigenvectors, $\lvert \mu_p\rangle$, allow
for a simple algorithm to construct the eigenvectors of SIS symmetrized
model, whose relevant $s=N/2$ eigensector dimension equals to
$N+1$. Therefore, investigation of finite symmetries brings down ABEM to
the same footing of compartmental models regarding the number of agents,
but does not neglect the role played by fluctuations.

\begin{acknowledgments}
 A.S.M. holds grants from CNPq 485155/2013 and
307948/2014-5, G.C.C. acknowledges funding from CAPES 067978/2014-01 and
A.C.P.M acknowledges grant CNPq 800585/2016-0.
\end{acknowledgments}

\appendix

\section{Algorithms}

\subsection{Time evolution}
\begin{algorithm}[H]
\caption{\label{algo1}}
\begin{algorithmic}[1]
\Require $p\in \mathbbm{N}$, matrix $A$ and off-diagonal transitions
\State $S=\{ \quad \}$\Comment{Basis}
\For {$\mu=0$ to $\mu<2^N$}
\State $\psi,\mathcal{N}_{\psi}\leftarrow \text{calculates eigenvector
  and norm from}\, \mu$
\State Add $\psi$ to $S$
\EndFor\Comment{$p$ invariant eigensector}
\For {$\psi$ in $S$}
\For {$k=0$ to $k < N$}
\State $\psi^{\prime}\leftarrow \text{off-diagonal transitions from $k$-th
  component of}\, \psi$
\State Evaluate $T_{\psi^{\prime}\psi}$ \Comment{Sparse storage}
\EndFor
\EndFor
\State $\pi\leftarrow$ initial condition
\For {$t=0$ to $t<t_{\text{max}}$}
\State $\pi\leftarrow \hat{T}\times \pi$
\EndFor\Comment{End time evolution}
\end{algorithmic}
\end{algorithm}

\subsection{Number of infected agents}

\begin{algorithm}[H]
\caption{\label{algo_count}}
\begin{algorithmic}[1]
\Function{count}{$\mu$,count}
\State c $\leftarrow \mu$
\State count $\leftarrow$ 0
\For{$k=0$ to $k<N$}
\State count $\leftarrow$ count + c \% 2
\State c $\leftarrow$ c // 2
\EndFor
\EndFunction
\end{algorithmic}
\end{algorithm}
\begin{Verbatim}[commandchars=\\\{\},codes={\catcode`\$=3\catcode`\^=7\catcode`\_=8}]
\PYG{k}{def} \PYG{n+nf}{count}\PYG{p}{(}\PYG{n}{mu}\PYG{p}{)}
  \PYG{n}{configuration}  \PYG{o}{=} \PYG{n}{mu}  \PYG{c+c1}{\PYGZsh{}$\mu = 0,1,\ldots,2^{N}-1$}
  \PYG{n}{count} \PYG{o}{=}\PYG{l+m+mi}{0}
  \PYG{k}{for} \PYG{n}{k} \PYG{o+ow}{in} \PYG{n+nb}{range}\PYG{p}{(}\PYG{n}{N}\PYG{p}{):}
    \PYG{n}{count} \PYG{o}{=} \PYG{n}{count} \PYG{o}{+} \PYG{p}{(}\PYG{n}{configuration} \PYG{o}{\PYGZpc{}} \PYG{l+m+mi}{2}\PYG{p}{)}
    \PYG{n}{conf}  \PYG{o}{=} \PYG{n}{conf} \PYG{o}{//} \PYG{l+m+mi}{2}
  \PYG{k}{return} \PYG{n}{count}
\end{Verbatim}
The symbol $\%$ stands for modulo integer operation while double
forward slashes stands for integer division. 

\subsection{Representative vectors}

\begin{algorithm}[H]
\caption{\label{algo_representative}}
\begin{algorithmic}[1]
\Function{representative}{$\mu,\psi,r$}
\State $\psi \leftarrow \mu$
\State $r \leftarrow 1$
\For{$k=0$ to $k < N-1$}
\State $\mu \leftarrow \hat{P} \mu$
\If{ $\mu < \psi $}
\State $\psi \leftarrow \mu$
\ElsIf{$\mu = \psi$}
\State $r\leftarrow r+1$
\EndIf
\EndFor
\EndFunction
\end{algorithmic}
\end{algorithm} 
\begin{Verbatim}[commandchars=\\\{\},codes={\catcode`\$=3\catcode`\^=7\catcode`\_=8}]
\PYG{k}{def} \PYG{n+nf}{get\PYGZus{}representative}\PYG{p}{(}\PYG{n}{mu}\PYG{p}{)}
  \PYG{n}{representative} \PYG{o}{=} \PYG{n}{mu}  \PYG{c+c1}{\PYGZsh{}inital guess for $\mu_0$}
  \PYG{n}{repetition}     \PYG{o}{=} \PYG{l+m+mi}{1}   \PYG{c+c1}{\PYGZsh{}initial repetition}
  \PYG{k}{for} \PYG{n}{k} \PYG{o+ow}{in} \PYG{n+nb}{range}\PYG{p}{(}\PYG{n}{N}\PYG{o}{\PYGZhy{}}\PYG{l+m+mi}{1}\PYG{p}{):}
    \PYG{n}{mu} \PYG{o}{=} \PYG{n}{permutation}\PYG{p}{(}\PYG{n}{mu}\PYG{p}{)} \PYG{c+c1}{\PYGZsh{}execute permutation over $\mu$}
    \PYG{k}{if} \PYG{o+ow}{not} \PYG{p}{(}\PYG{n}{mu} \PYG{o}{\PYGZgt{}} \PYG{n}{representative}\PYG{p}{):}
      \PYG{c+c1}{\PYGZsh{} update repetition}
      \PYG{n}{repetition} \PYG{o}{=} \PYG{n}{repetition} \PYG{o}{+} \PYG{n+nb}{max}\PYG{p}{(}\PYG{l+m+mi}{0}\PYG{p}{,} \PYG{l+m+mi}{1}\PYG{o}{\PYGZhy{}}\PYG{n+nb}{abs}\PYG{p}{(}\PYG{n}{representative}\PYG{o}{\PYGZhy{}}\PYG{n}{mu}\PYG{p}{))}
      \PYG{c+c1}{\PYGZsh{} if $\mu$ == $\mu_0$, max(0,1)=1; null otherwise}
      \PYG{n}{representative} \PYG{o}{=} \PYG{n}{mu} \PYG{c+c1}{\PYGZsh{}update representative $\mu_0$}
  \PYG{k}{return} \PYG{n}{representative}\PYG{p}{,}\PYG{n}{repetition}
\end{Verbatim}

\subsection{One-body off-diagonal transitions}
\begin{algorithm}[H]
\caption{\label{algo_onebody}}
\begin{algorithmic}[1]
  \Function{onebody}{label,rules,output}
  \For{$k=0$ to $k<N$}
  \Comment{Loop over agents}
  \If{ label[k] in rules}
  \State {new $\leftarrow$ label with label[k] $\leftarrow$ rule[label[k]][0]}
  \State output[new] $\leftarrow$ coupling 
  \EndIf
  \EndFor
  \EndFunction
\end{algorithmic}
\end{algorithm}
\begin{Verbatim}[commandchars=\\\{\},codes={\catcode`\$=3\catcode`\^=7\catcode`\_=8}]
\PYG{k}{def} \PYG{n+nf}{onebody}\PYG{p}{(}\PYG{n}{conf}\PYG{p}{,}\PYG{n}{output}\PYG{o}{=}\PYG{p}{\PYGZob{}\PYGZcb{}):}
  \PYG{k}{for} \PYG{n}{k} \PYG{o+ow}{in} \PYG{n+nb}{range}\PYG{p}{(}\PYG{n}{N}\PYG{p}{):}     \PYG{c+c1}{\PYGZsh{}agent loop}
    \PYG{k}{if} \PYG{n}{conf}\PYG{p}{[}\PYG{n}{k}\PYG{p}{]} \PYG{o+ow}{in} \PYG{n}{rules}\PYG{p}{:} \PYG{c+c1}{\PYGZsh{}check for k\PYGZhy{}th agent transition}
      \PYG{c+c1}{\PYGZsh{}new configuration}
      \PYG{c+c1}{\PYGZsh{}    rules == dictionary}
      \PYG{c+c1}{\PYGZsh{}    rules[key] == (new state, coupling)}
      \PYG{n}{new}\PYG{o}{=}\PYG{n}{conf}\PYG{p}{[:}\PYG{n}{k}\PYG{p}{]}\PYG{o}{+}\PYG{n}{rules}\PYG{p}{[}\PYG{n}{conf}\PYG{p}{[}\PYG{n}{k}\PYG{p}{]][}\PYG{l+m+mi}{0}\PYG{p}{]}\PYG{o}{+}\PYG{n}{conf}\PYG{p}{[}\PYG{n}{k}\PYG{o}{+}\PYG{l+m+mi}{1}\PYG{p}{:]}
      \PYG{c+c1}{\PYGZsh{}output: new == key, coupling == value}
      \PYG{n}{output}\PYG{p}{[}\PYG{n}{new}\PYG{p}{]}\PYG{o}{=}\PYG{n}{rules}\PYG{p}{[}\PYG{n}{conf}\PYG{p}{[}\PYG{n}{k}\PYG{p}{]][}\PYG{l+m+mi}{1}\PYG{p}{]}
  \PYG{k}{return} \PYG{n}{output}
\end{Verbatim}

\subsection{Two-body off-diagonal transitions}
\begin{algorithm}[H]
\caption{\label{algo_twobody}}
\begin{algorithmic}[1]
\Function{twobody}{L,A,rules,outcome}
\For{$j=0$ to $j < N$}
\For{$i=0$ to $i < N$}
\State $q\leftarrow (L_jL_i)$
\If{ $q$ in rules}
\State $x \leftarrow$ L
\State $x_j\leftarrow \text{rules[q]}_{00}$
\State $x_i\leftarrow \text{rules[q]}_{01}$
\State $\text{output}[x] \leftarrow \text{output}[x]+A_{j i}$
\EndIf
\EndFor
\EndFor
\EndFunction
\end{algorithmic}
\end{algorithm}
\begin{Verbatim}[commandchars=\\\{\},codes={\catcode`\$=3\catcode`\^=7\catcode`\_=8}]
\PYG{k}{def} \PYG{n+nf}{twobody}\PYG{p}{(}\PYG{n}{conf}\PYG{p}{,}\PYG{n}{A}\PYG{p}{,}\PYG{n}{output}\PYG{o}{=}\PYG{p}{\PYGZob{}\PYGZcb{}):}
  \PYG{k}{for} \PYG{n}{j} \PYG{o+ow}{in} \PYG{n+nb}{range}\PYG{p}{(}\PYG{n}{N}\PYG{p}{):}
    \PYG{n}{sj}\PYG{o}{=}\PYG{n}{conf}\PYG{p}{[}\PYG{n}{j}\PYG{p}{]}
    \PYG{n}{trialj}\PYG{o}{=}\PYG{n}{conf}\PYG{p}{[:}\PYG{n}{j}\PYG{p}{]}
    \PYG{k}{for} \PYG{n}{i} \PYG{o+ow}{in} \PYG{n+nb}{range}\PYG{p}{((}\PYG{n}{j}\PYG{o}{+}\PYG{l+m+mi}{1}\PYG{p}{),}\PYG{n}{N}\PYG{p}{):}
      \PYG{n}{pair}\PYG{o}{=} \PYG{n}{sj}\PYG{o}{+}\PYG{n}{conf}\PYG{p}{[}\PYG{n}{i}\PYG{p}{]}
      \PYG{k}{if} \PYG{n}{pair} \PYG{o+ow}{in} \PYG{n}{rules}\PYG{p}{:}
        \PYG{n}{out}\PYG{o}{=}\PYG{n}{rules}\PYG{p}{[}\PYG{n}{pair}\PYG{p}{]}
        \PYG{c+c1}{\PYGZsh{} new configuration}
        \PYG{n}{trial}\PYG{o}{=}\PYG{n}{conf}
        \PYG{n}{trial}\PYG{p}{[}\PYG{n}{j}\PYG{p}{]}\PYG{o}{=}\PYG{n}{out}\PYG{p}{[}\PYG{l+m+mi}{0}\PYG{p}{][}\PYG{l+m+mi}{0}\PYG{p}{]}
        \PYG{n}{trial}\PYG{p}{[}\PYG{n}{i}\PYG{p}{]}\PYG{o}{=}\PYG{n}{out}\PYG{p}{[}\PYG{l+m+mi}{0}\PYG{p}{][}\PYG{l+m+mi}{1}\PYG{p}{]}
        \PYG{k}{if} \PYG{n}{trial} \PYG{o+ow}{not} \PYG{o+ow}{in} \PYG{n}{output}\PYG{p}{:}
          \PYG{n}{output}\PYG{p}{[}\PYG{n}{trial}\PYG{p}{]}\PYG{o}{=}\PYG{l+m+mi}{0}
        \PYG{c+c1}{\PYGZsh{} update coupling}
        \PYG{n}{output}\PYG{p}{[}\PYG{n}{trial}\PYG{p}{]} \PYG{o}{+=} \PYG{n}{out}\PYG{p}{[}\PYG{l+m+mi}{1}\PYG{p}{]}\PYG{o}{*}\PYG{n}{A}\PYG{p}{[}\PYG{n}{j}\PYG{p}{][}\PYG{n}{i}\PYG{p}{]}
  \PYG{k}{return} \PYG{n}{output}
\end{Verbatim}

%

\end{document}